\documentclass[a4, 11pt]{article}
\usepackage[textwidth=16cm]{geometry}
\usepackage{amssymb, amsmath}
\usepackage{graphicx}
\usepackage{color,cite,authblk}

\newcommand{\fig}[1]{Fig.~\ref{#1}}
\newcommand{\eq}[1]{Eq.~(\ref{#1})}

\newcommand{\dt}{\partial_{t}}
\newcommand{\dxs}{\partial_{x}}
\newcommand{\dx}{\partial_{\mathbf{x}}}

\newcommand{\vecx}{\mathbf{x}}

\newcommand{\mat}[4]{\left(\begin{array}{cc}
                   #1 & #2\\
                   #3 & #4
                 \end{array} \right)}

\begin{document}
\title{Population extinction in an inhomogeneous host-pathogen model}
\author{Trilochan Bagarti\thanks{bagarti@hri.res.in}}
\affil{Harish-Chandra Research Institute\\ Chhatnag Road, Jhunsi, Allahabad-211019}

\maketitle
\begin{abstract}
We study inhomogeneous host-pathogen dynamics to model the global amphibian population extinction in a lake basin system. The lake basin system is modeled as quenched disorder. In this model we show that once the pathogen arrives at the lake basin it spreads from one lake to another, eventually spreading to the entire lake basin system in a wave like pattern. The extinction time has been found to depend on the steady state host population and pathogen growth rate. Linear estimate of the extinction time is computed. The steady state host population shows a threshold behavior in the interaction strength for a given growth rate.
\end{abstract}

\section{Introduction}
The problem of population extinction in ecological systems has been studied extensively both theoretically and experimentally. Population extinction can occur due to a number of reasons such as habitat loss, climate change, pollution, epidemics, etc\cite{szwa,erik,gott,poun,vred,craw}. Also, when a species alien to a stable ecological system is introduced it may pose a threat of extinction for a certain number of species\cite{sher, will, sato}.

In an ecological system the underlying dynamics could be quite complex due to its large degrees of freedom\cite{rjwill}. A theoretical model is often helpful in understanding the dynamics of extinction without going into finer details of the real system. One approach would be to include in the model only the relevant degrees of freedom and incorporate the effect of the rest in parameters that can be determined experimentally. Population model with spatial inhomogeneity have been studied by a number of authors\cite{murray1,cant,dahm2,meer2,miss,miss2,kuma,joo}. Modelling extinction in inhomogeneous system can further our understanding of the dynamics of real systems. For instance spatial inhomogeneity has a stabilizing effect on the interaction among different species\cite{comi,whit}, it can affect the front speed\cite{kine} and the width\cite{hecht} of a propagating invasion. 

The simplest example of inhomogeneous reaction-diffusion system is the trapping of diffusing particle by traps\cite{havl}. The density of particles for trapping reaction with randomly distributed traps has been know to show a stretched exponential behavior\cite{balagurov,donsker,gras}, self-segregation\cite{jayan1,wio,htait,koza}, and  self-organization around traps\cite{wei89,tait,jayan2,thm}. Random growth and trapping of diffusing particle can be used to model inhomogeneous population models\cite{jayan3,krpv}. Recently trapping reaction models has been used to study population dynamics with localized predation\cite{tab}.

In this paper we study host-pathogen dynamics in a spatially inhomogeneous system. Our motivation come from the recent studies of global amphibian population extinction due to infection by the fungal pathogen {\it Batrachochytrium dendrobatidis}(Bd)\cite{vred, craw}. Detailed investigation of the frog-Bd dynamics in lake basins have shown that after the arrival of the pathogen in a basin it spreads in a wave like pattern infecting the entire population. It is likely that the spread of pathogen occurs via an unknown vector since the inter-lake frog movement has not been found. Complete extinction is observed only when the intensity of infection reached a critical threshold \cite{vred}. Here we study a model of host-pathogen dynamics that explains these observed phenomena qualitatively. We shall show how the pathogen spreads from one lake to another by a wave like pattern and estimate the linear wave speed. The extinction time at a lake depends on the steady state population of a lake. We compute the extinction time analytically and numerically. By numerical computations we shall investigate the host-pathogen dynamics in a lake basin system to understand global population decline. Further applications and improvements to the model are discussed.

\section{Formulation of the model}
\label{react.diff}

We consider interaction between fungal pathogen `Bd' and host `frog' in an inhomogeneous media. The inhomogeneous media consists of localized regions in space that are habitats of frogs. These localized regions for our case models the lake basin systems described in Ref.\cite{vred}. As the interlake movement of frogs has not been observed, we assume that the pathogen interacts with the host only in these localized regions. The size of lakes are assumed negligible when compared to the mean separation between two nearest lakes. Therefore, we assume the lakes to be point like regions in space, as a result host-pathogen dynamics in the lake basin system reduces to a host-pathogen model in presence of quenched disorder\cite{havl}. This assumption is valid as long as we are interested in the depletion of the total host population in the lake basin. However, a generalization is possible where one can replace the point objects by extended objects. Such a model would be interesting only when one is interested in intra-lake dynamics with lakes of various shapes and sizes.

Let us denote by $b(\vecx,t)$ the density of pathogen at position $\vecx$ at time $t$ and the host population in the $i$-th lake by $f_i(t)$. Reaction-diffusion equation for the host-pathogen dynamics is given by
\begin{align}
\dt b(\vecx,t) &= D\dx^2 b(\vecx,t) + \Phi(\vecx) b(\vecx,t)-\mu b(\vecx,t) + \frac{\epsilon b(\vecx,t)}{1+b(\vecx,t)/b_0},\nonumber \\
\dot{f}_i(t) &= \gamma\left[1-\theta_i f_i(t)\right]f_i(t) - \alpha b(\vecx_i, t)f_i(t),~~~i=1,2,\ldots,
\end{align}
where $\dot{}=d/dt$, $D$ is the diffusion constant of pathogen, the function $\Phi(\vecx) = \rho \sum_i\delta(\vecx-\vecx_i) f_i(t)$ is the spatially varying host dependent growth rate of the pathogen, $\mu$ is the pathogen decay rate and the term $\epsilon b/(1+b/b_0)$ describes the growth of pathogen in the absence of host with $\epsilon$ and $b_0$ constants. Note that the choice of this particular form is to ensure a bounded growth of the pathogen population. At low densities the pathogen growth is exponential where as at high density it is constant. Such growth is seen when there is a competition of resources. The function $\Phi(\vecx)$ models the lakes as point like object in space with the $i$th lake located at position $\vecx = \vecx_i$.  The boundary condition is given by $\lim_{|\vecx|\rightarrow \infty} b(x,t) = 0$ and initial conditions $b(x,0) = \delta(\vecx-\vecx_1)$, $f_i(0)=1$ for all $i=1,2,\ldots,N$.

The host population $f_i$ obeys a logistic equation with a growth rate $\gamma(1-\theta_i f_i)$ and the interaction term $-\alpha b(\vecx_i,t)f_i$ describes the decay of host population due to pathogen infection. Note that the decoupled equations (i.e. for $\rho = \alpha = 0$) have homogeneous steady states $b^*=0,~\epsilon/\mu-1$ and $f_i^* = 0,~1/\theta_i$. This implies that host and the pathogen can sustain their populations separately provided the nonzero steady states are stable. Let $t \rightarrow Dt,~\rho \rightarrow \rho/D,~\mu \rightarrow \mu/D,~\epsilon \rightarrow \epsilon/D,~\gamma \rightarrow \gamma/D,\alpha \rightarrow b_0\alpha/D$  and $~b \rightarrow b/b_0$ so that we have
\begin{align}
\dt b(\vecx,t) &= \dx^2 b(\vecx,t) + \Phi(\vecx) b(\vecx,t) - \mu b(\vecx,t) + \frac{\epsilon b(\vecx,t)}{1+b(\vecx,t)},\nonumber \\
\dot{f}_i(t) &= \gamma\left[1-\theta_i f_i(t)\right]f_i(t) - \alpha b(\vecx_i, t)f_i(t),~~~i=1,2,\ldots,
\label{diffeq2}
\end{align}
with the boundary condition $\lim_{|x|\rightarrow \infty} b(x,t)=0$ and initial condition $b(x,0)=\delta(\vecx-\vecx_1)/b_0$, $f_i(0)=1$ for all $i=1,2,\ldots,N$.
\section{Steady state and linear stability}
Let us consider the homogeneous steady states $b^*=0,~\epsilon/\mu-1$ and $f_i^* = 0,~1/\theta_i$ of the decoupled equations. For the host, the homogeneous steady state $f_i^*=0$ is unstable and $f_i^*=1/\theta_i$ is stable. In other words, without the influence of pathogen each lake can be considered as a stable ecosystems. For the pathogen, let $b-b^* \sim \exp(i\mathbf{k x} + \omega t)$ be as small perturbation so that \eq{diffeq2} gives
\begin{equation}
\omega = -|\mathbf{k}|^2 - \mu +\epsilon/(1+b^*).
\end{equation}
Clearly, the steady state $b^*$ is stable if $\epsilon/(1+b^*) \leq \mu$. This implies that the steady state $b^*=0$ is stable if $\epsilon<\mu$. However, if $\epsilon > \mu$ there exist a range of unstable modes $\mathbf{k}$ such that $-\sqrt{\epsilon-\mu} < |\mathbf{k}| < \sqrt{\epsilon-\mu}$ for which which $b^*=0$ becomes unstable. On the other hand, the steady state $b^*=\epsilon/\mu-1$ is stable for $\epsilon > \mu$. 

One would be interested to know how these steady states change when host-pathogen interaction is introduced. We expect the host population to decline due to the pathogen interaction. Let us assume that a small patch of pathogen have just arrived at a particular lake $i$ at time $t=0$. We want to understand how this small patch grow in time. We shall estimate the time scale for the total extinction of the host in the lake. 

\begin{figure}
\centering
\includegraphics[width=0.6\textwidth]{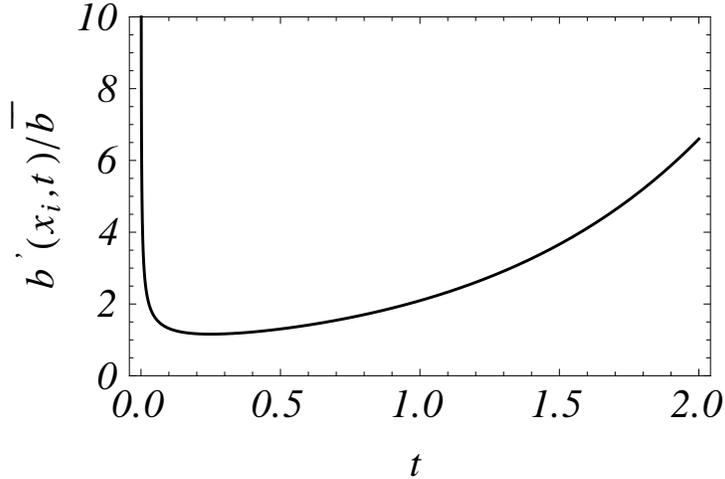}
\caption{Growth of pathogen population in the $i$th lake. At $t=0$ the lake is infected by introducing a small perturbation to the steady state. Initially it decreases as $t^{-1/2}$ then there is an exponential growth. Parameters: $\rho/\theta_i =1, \epsilon-\mu=1.$}
\label{linbgrw}
\end{figure}

Assume that the lake $i$ has a steady state population $f_i(t)=1/\theta_i$ for $t \leq 0$ and at time $t=0$ we introduce a small perturbation $b'(x,0)= \delta(\vecx-\vecx_i)\bar{b}$ to the steady state $b^*=0$ where $0 < \bar{b} \ll 1$. Linearizing \eq{diffeq2} near $b^*=0$ and $f_i^*=1/\theta_i$ we obtain
\begin{align}
\dt b'(\vecx,t) &= \dx^2 b'(\vecx,t) + \rho \theta_i^{-1}\delta(\vecx-\vecx_i) b'(\vecx,t) +(\epsilon - \mu) b'(\vecx,t),\nonumber \\
\dot{f'}_i(t) &= -\gamma f'_i(t) - \alpha \theta_i^{-1} b'(\vecx_i, t),
\label{decoup.diffeq} 
\end{align}
with boundary conditions $b'=0$ as $|\mathbf{x}|\rightarrow \infty$ and initial conditions $b'(x,0) = \delta(\vecx-\vecx_i)\bar{b}$, $f'_i(0)=0$. Note that in \eq{decoup.diffeq} we have ignored all lakes other than the lake $i$ as we are interested only in the local dynamics. We notice from \eq{decoup.diffeq} that the equation for the pathogen can be solved exactly and the solution can then be used to obtain the host population $f_i$. In order to find a solution $b'$ let us transform $b' \rightarrow b'\exp(-(\epsilon-\mu)t)$ so that we have the simplified equation $\dt b'(\vecx,t) = \dx^2 b'(\vecx,t) + \rho \theta_i^{-1}\delta(\vecx-\vecx_i) b'(\vecx,t)$. Taking Laplace transform we obtain the solution
\begin{equation}
b'(\vecx_i,s) = \frac{\bar{b}G(\vecx_i|\vecx_i)}{1-\theta^{-1}_i \rho G(\vecx_i|\vecx_i)},
\end{equation}
where $G(\vecx|\mathbf{y})$ denotes the Green's function defined by $G=(s-\dx^2)^{-1}$. Substituting for the Green's function, $G(\vecx|\mathbf{y}) = \exp(-\sqrt{s}|\vecx-\mathbf{y}|)/2\sqrt{s}$, taking inverse Laplace transform and multiplying the factor $\exp((\epsilon-\mu)t)$ we obtain
\begin{equation}
b'(\vecx_i,t)= \bar{b}~\mbox{e}^{(\epsilon-\mu) t}\left(\frac{1}{\sqrt{4 \pi t}} + \frac{\rho}{4\theta_i}\mbox{e}^{\rho^2 t/(4\theta^2_i)} \mbox{erfc}(-\frac{\rho t^{1/2}}{2\theta_i})\right).
\label{linb}
\end{equation}
In \fig{linbgrw} we plot the ratio $b'(\vecx_i,t)/\bar{b}$ as a function of time. The divergence at $t=0$ is due to the initial condition we have chosen. Initially there is a drop in the population $b'$ due to the $t^{-1/2}$ term that arises as a result of pathogen diffusing out of the lake. We observe that the population grows exponentially at a rate $K = \rho^2/4\theta_i^2$. The fact that $K$ is proportional to the square of steady state host population $1/\theta_i$ indicates that pathogen growth rates in thickly populated lakes will be large. 

\begin{figure}
\centering
\includegraphics[width=0.6\textwidth]{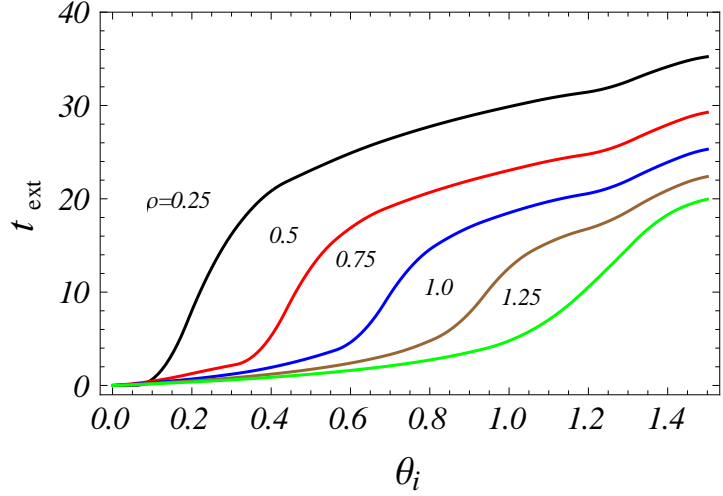}
\caption{Extinction time $t_{\text{ext}}$ as a function of $\theta_i$. Parameters: $D=0.1,~\gamma=1,~\alpha=0.5,~\mu=0.25,~\epsilon=0.8, ~\bar{b}=0.1$.}
\label{ext.time}
\end{figure}

\begin{figure}
\centering
\includegraphics[width=0.6\textwidth]{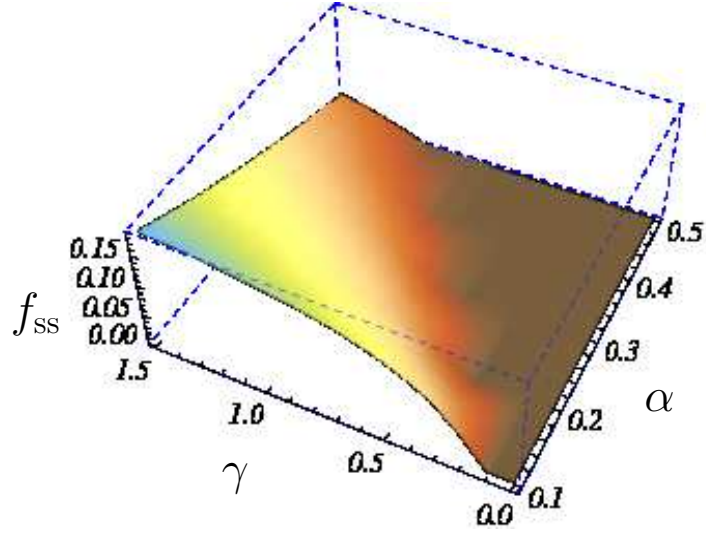}
\caption{Steady-state host population $f_{\text{ss}}$ as a function of growth $\gamma$ and pathogen interaction strength $\alpha$. Parameters: $D=0.1,~\rho=1.0,\mu=0.25,~\epsilon=0.8,~\theta=1.0$.}
\label{steadystate.popl}
\end{figure}

The host population $f_i(t)$ can be written as
\begin{equation}
f_i(t) = \frac{1}{\theta_i}-\frac{\alpha}{\theta_i}\int_0^t b'(\vecx_i,\tau)\mbox{e}^{-\gamma(t-\tau)}d\tau
\label{linf}
\end{equation}
Substituting the expression for $b'(\vecx_i,t)$ in \eq{linf} we obtain 
\begin{equation}
f_i(t) = \frac{1}{\theta_i} - \frac{\alpha \bar{b}\mbox{e}^{-\gamma t}}{2\theta_i(K+R)}\left\{\sqrt{K}\left(\mbox{e}^{(K+R)t}\left[1+\mbox{erf}(\sqrt{Kt})\right]-1\right)+ \sqrt{R}~\mbox{erfi}(\sqrt{Rt})\right\},
\label{linf2}
\end{equation}
with $R=\epsilon-\mu+\gamma$. Now we can estimate the time scale $t_{\text{ext}}$ for the complete extinction of the host population by solving the equation $f_i(t)=0$. Taking only the fastest growing term in \eq{linf2} we obtain
\begin{equation}
t_{\text{ext}} \simeq \frac{1}{K+R-\gamma}\log \left(\frac{4(K+R)}{3\bar{b}\alpha\sqrt{K}}\right).
\label{hostdec}
\end{equation}
Note that to obtain \eq{hostdec} the $\mbox{erf}(\sqrt{Kt})$ has been replaced by its mean. Expression for $t_{\text{ext}}$ in the original variables can be obtained by inverting the transformation we have used. The actual time for extinction shall however be larger than that we have obtained in \eq{hostdec} from the linearized equations. The linear estimate provides only a clue as to how it will depend on the parameters. For example, the extinction time $t_{\text{ext}}$ in \eq{hostdec} shows a sigmoidal behavior $\theta_i$. This implies that lakes that are sparsely populated will take a very long time for extinction. On the other hand for thickly populated lakes it is the opposite. In \fig{ext.time} we have computed numerically the extinction time for different values of $\rho$. We define extinction time $t_{\text{ext}}$ to be the time required for the host population to reduce to a fraction $f$ of it initial population. Here we assume $f = 0.01$. We note that for $\theta_i$ close to zero i.e. for thickly populated lake the extinction time is very small. As $\theta_i$ increases it rapidly grows for a small range of $\theta_i$ and then shows saturation. The numerical values obtained from \eq{hostdec} is approximately one order of magnitude less. This is expected since the linearized equation assumes a constant pathogen growth rate $K$. We know that the pathogen growth rate decreases as the host population decreases therefore the linear estimates should be regarded as the lower bound of the extinction time.
The steady state host population $f_{\text{ss}}$ as a function of growth rate $\gamma$ and interaction strength $\alpha$ is shown in \fig{steadystate.popl}. For $\alpha =0$ we know that $f_{\text{ss}}=1/\theta_i$ which is independent of the growth rate $\gamma$. When there is host-pathogen interaction i.e. $\alpha \neq 0$, the steady-state changes. We choose a range of values of $\alpha$ and $\gamma$ keeping other parameters constant. We observe that for a given value of growth rate $\gamma$ there exists a critical value $\alpha_c$ such that $f_{\text{ss}}$ vanishes for all $\alpha>\alpha_c$. In the real situation each lake can have a different growth rate $\gamma_i$ and interaction strength $\alpha_i$ which can give rise to more complex phenomena.

\section{Host-pathogen dynamics in a two lake system}
To see how the pathogen spreads let us consider a two lake system in one spatial dimension with the lakes located at points $x=0$ and $x=L$. At time $t=0$ we assume that the lake at the origin is infected and the lake at $x=L$ is free of pathogen. The pathogen spreads into the neighborhood of the lake by a travelling wave. To determine the speed of the travelling wave let us consider the one dimensional case without interaction i.e. $\Phi = 0$. We have 
\begin{equation}
\dt b= \dxs^2 b - \mu b + \epsilon b/(1+b).
\label{traveleqn}
\end{equation}
A travelling wave solution $b(x,t)=u(x-ct)$ satisfies the \eq{traveleqn} so that in a moving frame we can write
\begin{equation}
\frac{d^2u}{dz} +c \frac{du}{dz} +\left(\frac{\epsilon}{1+u}-\mu\right)u= 0,
\label{traveleqn2}
\end{equation}
with $z = x - ct$ and boundary conditions 
\begin{equation}
\lim_{z \rightarrow -\infty} u(z) = \epsilon/\mu-1 \mbox{~~and~~} \lim_{z \rightarrow \infty} u(z) = 0,
\end{equation}
where $u= \epsilon/\mu-1 \mbox{~~and~~} 0$ are the homogeneous steady states of \eq{traveleqn}. The constant $c$ is the speed of the travelling wave. Defining $v = du/dz$ we can rewrite \eq{traveleqn2} as
\begin{equation}
\frac{du}{dz}=v,~~~\frac{dv}{dz} =-c v -\left(\frac{\epsilon}{1+u}-\mu\right)u= 0.
\label{phsp}
\end{equation}
The steady state solution of \eq{phsp} are $(u^*,v^*) = (0,0)$ and $(\epsilon/\mu-1,0)$.
Linearizing around the steady states we can write $dX/dz = MX$ where $X = (u'~v')^{T}$ denotes a perturbation to $(u^*~v^*)^{T}$ and the matrix 
\begin{equation}
M = \mat{0}{1}{\frac{\epsilon u^*}{(1+u^*)^2}-\frac{\epsilon}{1+u^*}+\mu}{-c}.
\end{equation}
The eigenvalues of $M$ characterizes the dynamics near the steady state $(u^*,v^*)$. The eigenvalues are
\begin{align}
\lambda &= -c/2 \pm \sqrt{\mu-\epsilon+c^2/4}, \mbox{~~~~~~for~} (0,0),\nonumber \\
\lambda &= -c/2 \pm \sqrt{(\epsilon-\mu)\mu/\epsilon+c^2/4}, \mbox{~~~~~for~} (\epsilon/\mu-1,0).
\end{align}
We note that $(0,0)$ is a stable node if $c \geq 2\sqrt{\epsilon-\mu}=c_{\text{min}}$ otherwise it is a stable spiral. The steady state $(\epsilon/\mu-1,0)$ is a saddle node. Travelling wave moving at speed less then $c_{\text{min}}$ are unphysical\cite{murray1}. 

\begin{figure}
\centering
\includegraphics[width=0.6\textwidth]{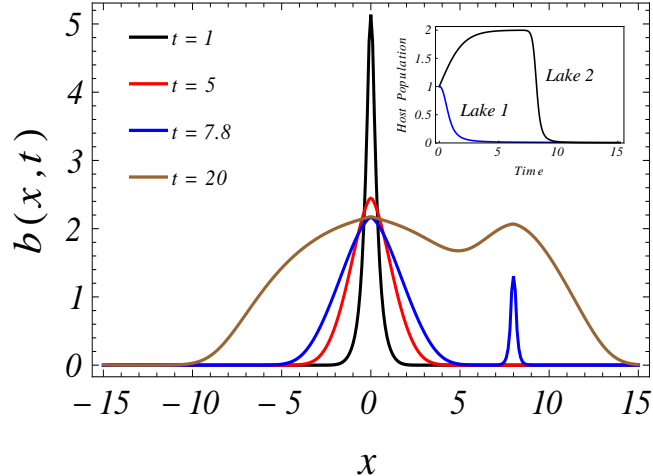}
\caption{Host-pathogen dynamics in a two lake system. Here Lake 1 and Lake 2 are the lake at the origin and at $x=L$ respectively Parameters: $D=0.1,~\epsilon=0.8,~\mu=0.25,\alpha=0.5,~\rho = 1,~\gamma = 1,~\theta = 0.5$.}
\label{ext.popl.1d}
\end{figure}

Now let us consider the two lake system. The lake at the origin is infected at time $t=0$ where as the lake at $x=L$ is free of pathogen. The time required for the pathogen to reach the lake is $t_{\text{inf}} \leq L/(2\sqrt{\epsilon-\mu})$. In \fig{ext.popl.1d} we have plotted the pathogen density for the two lake system. The lake 1 is located at the origin and lake 2 at $x=8$. Initially only lake 1 is infected by the pathogen so we have taken a delta function initial condition centered at the origin. As the system evolves in time the pathogen population spreads into the neighboring region. At time $t=7.8$ we see a small peak at lake 2 this is approximately the time when the lake is infected. The density of pathogen in the region between the lakes is negligible and the peak at lake 2 indicates that there is a sharp growth of pathogen after it has reached the lake. The host population therefore decreases exponentially (see \fig{ext.popl.1d} inset) due to the infection. Substituting the values $D=0.1,~\epsilon=0.8,~\mu=0.25,L=8$ we obtain $L/c_{\text{min}} = 17.05$ which is approximately twice of $t_{\text{inf}}$ obtained numerically. 
\section{Numerical result for multiple lakes}
\begin{figure}
\centering
\includegraphics[width=0.8\textwidth]{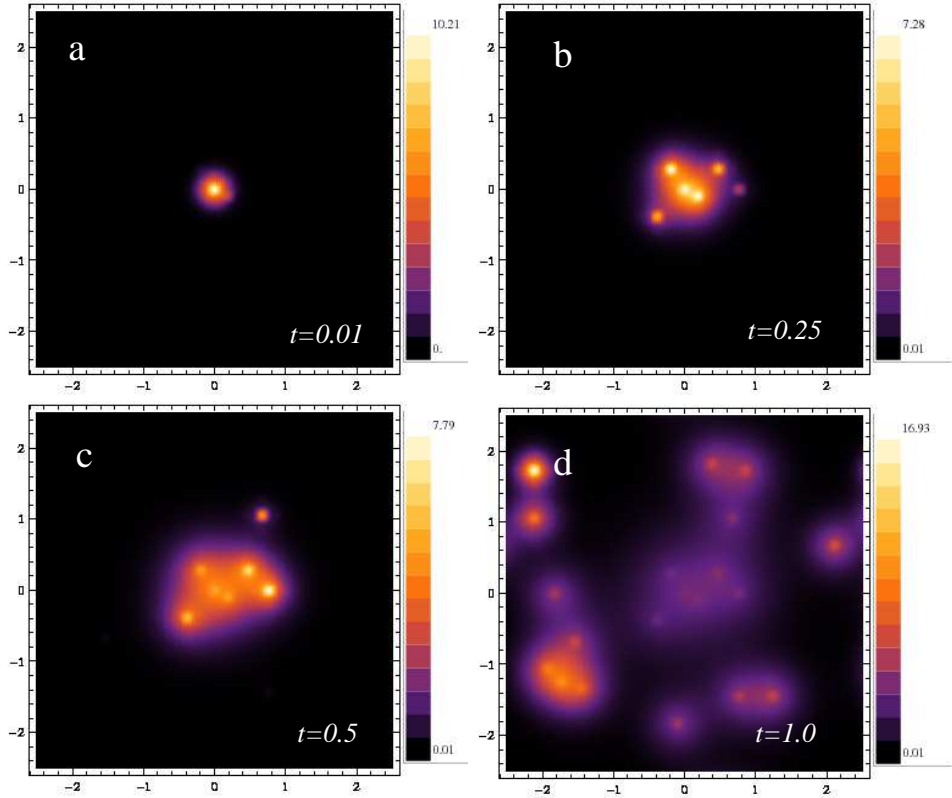}
\caption{Pathogen density shows wave like pattern in a multiple lake system. Here lake 1 is located at the origin which is infected at $t=0$ and lake 2-$N$ with $N=20$ are located at random positions. Parameters: $D=0.25, \epsilon=0.8, \mu=0.5, \rho=1.0, \gamma=1, \alpha=0.5$ and $\theta_i=0.1$ for all $i=1,\ldots N$.}
\label{popl.2d}
\end{figure}

\begin{figure}
\centering
\includegraphics[width=0.6\textwidth]{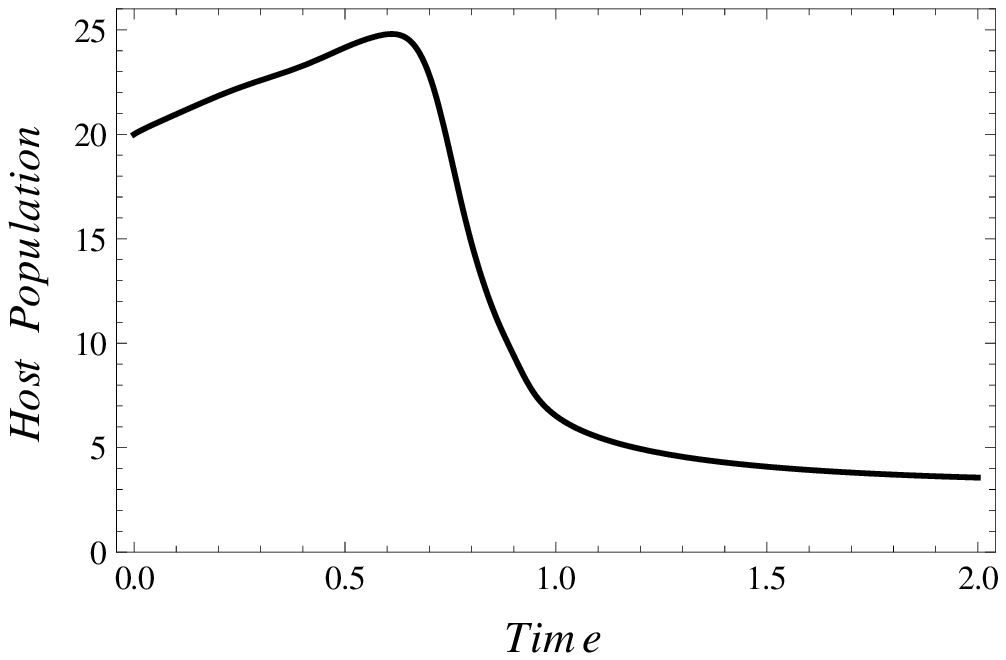}
\caption{Global population decline in a multiple lake system. Parameters: $D=0.25, \epsilon=0.8, \mu=0.5, \rho=1.0, \gamma=1, \alpha=0.5$ and $\theta_i=0.1$ for all $i=1,\ldots N$.}
\label{global.popl}
\end{figure}

Numerical computation for the lake basin system in two dimensions is done by finite difference method. The lake basin system consists of $N$ lakes at random positions $\vecx_1,\ldots,\vecx_N$ that are drawn from a uniform distribution. For our case we take $N=20$ and assume $\vecx_1$ as the origin which is the lake infected at time $t=0$. Initial host population $f_i(0) =1$ for all $i=1,\ldots,N$. We assume $\theta_i=0.1$ for all $1 \leq i \leq 20$ and $\rho=\gamma=1$. The parameters $\epsilon=0.8,~\mu=0.5$ and $\alpha=0.5$ are chosen. As larger values of diffusion constant flattens the density profile $b(\vecx,t)$ we choose as small $D=0.25$ for which the travelling wave is clearly observable. In \fig{popl.2d}(a) we can see the density $b(x,t)$ at $t=0.01$ which shows a peak at the origin i.e.. lake 1. As time progresses the pathogen spreads out infecting the neighboring lakes. The infection spreads from one lake to the nearest neighboring lake. The new lakes that are infected infected acts as new source of pathogen which then infect their neighboring lakes. The pathogen population grows as long as the lakes are populated by the host. After complete extinction at a lake the pathogen population reaches it steady state. In \fig{popl.2d}(d) we observe that pathogen density close to the origin has reduced as compared to \fig{popl.2d}(b). We also observe that new lakes that are further away from the lakes at the origin has been infected. A wave like pattern can be seen clearly from \fig{popl.2d}(a-d). We have computed the global host population as a function of time which is plotted in \fig{global.popl}. We observe that the global population decline qualitatively agrees with the experimental observations. We would like to emphasize here that quantitative prediction of the model can be tested only when growth and decay rates of the pathogen and the interaction strengths are provided. 

\section{Concluding remarks}
We studied an inhomogeneous host-pathogen model to understand the global amphibian population extinction. We assumed that the pathogen has a bounded growth and a linear decay. The host population on the other hand is confined to the lakes and is assumed to have logistic growth. Inhomogeneity is introduced as quenched disorder to model the lake basin system. The host-pathogen interaction is assumed bilinear in host population and the pathogen density. With this minimal model we have shown qualitatively the phenomena that has been observed experimentally. We observed that the pathogen spreads in a wave like pattern infecting one lake after another and eventually spreading to the entire lake basin. By linear analysis we obtained the extinction time and minimum wave speed which suggests that lakes that are thickly populated become extinct at a higher rate. The extinction rate is proportional to the square of the steady state host population. We calculated a lower bound of the extinction time.

A natural response of the system to the pathogen invasion should be the host migration. It would be interesting to see how extinction could be suppressed by including host migration. Also, to include intra lake dynamics the point like objects can be replaced by extended objects to model lakes of different shapes and sizes. The parameters needed in this model may be obtained locally from a sample of lakes assuming that all the lakes provide similar habitat for the host. Furthermore, for quantitative predictions parameters such as diffusion constant and growth(decay) rates should be determined from experiments.


\begin{thebibliography}{13}
\bibitem{szwa}J.~Szwabinski and A.~Pekalski, {Physica A}, {\bf 360}, 59(2006).
\bibitem{erik}A.~Eriksson, F. Elias-Wolff, and B. Mehlig, {Theo. Popl. Biol.}, {\bf 83}, 101(2013).
\bibitem{gott}O.~Gottesman and B.~Meerson, {Phys. Rev. E}, {\bf 83}, 021140(2012).
\bibitem{poun}J.~A.~Pounds, M.~R.~Bustamente, L.~A.~Coloma, J.~A.~Consuegra, et al. Nature {\bf 439}, 161(2006).
\bibitem{vred}V.~T.~Vredenburg, R.~A.~Knapp, T.~S.~Tunstall, and C.~J.~Briggs, {Proc. Natl. Acad. Sci. USA}, {\bf 107}, 9689(2010).
\bibitem{craw}A.~J.~Crawford, K.~R.~Lips, and E.~Bermingham, {Proc. Natl. Acad. Sci. USA}, {\bf 107}, 1377(2010).

\bibitem{sher}J.~A.~Sherratt, M.~A.~Lewis, and C.~Fowler, {Proc. Natl. Acad. Sci. USA}, {\bf 92}, 2524(1995).
\bibitem{will}J.~F.~Willemsen, {Physica A}, {\bf 389}, 3484(2010).
\bibitem{sato}K.~Sato, H. Matsuda and A.~Sasaki, {J. Math. Biol.}, {\bf 32}, 251(1994).

\bibitem{rjwill}R.~J.~Williams and N.~D.~Martinez, {Nature}, {\bf 404}, 180(2000).

\bibitem{murray1}J.~D.~Murray, Mathematical Biology, Ed. Second, Springer(1989).
\bibitem{cant}R.~S.~Cantrell and C. Cosner, {Spatial Ecology via Reaction-Diffusion Equations}, John Wiley \& Sons Ltd, England(2003).
\bibitem{dahm2}K.~A.~Dahmen, D.~R.~Nelson, and N.~M.~Shnerb, {J. Math. Biol.}, {\bf 41}, 1(2000).
\bibitem{meer2}B.~Meerson and P.~V.~Sasorov, {Phys. Rev. E}, {\bf 83}, 011129(2011).


\bibitem{miss}A.~R.~Missel and K.~A.~Dahmen, {Phys. Rev. Lett.}, {\bf 100}, 058301(2008).
\bibitem{miss2}A.~R.~Missel and K.~A.~Dahmen, {Phys. Rev. E}, {\bf 79}, 021126(2009).
\bibitem{kuma}N.~Kumar and V.~M.~Kenkre, {Physica A}, {\bf 390}, 257(2011).
\bibitem{joo}J.~Joo and J.~L.~Lebowitz, {Phys. Rev. E}, {\bf 72}, 036112(2005).

\bibitem{comi}H.~N.~Comins and D.~W.~E.~Blatt, {J. Theo. Biol.}, {\bf 48}, 75(1974).
\bibitem{whit}A.~White, M.~Begon, and R.~G.~Bowers, {Proc. R. Soc. Lond. B}, {\bf 263}, 325(1996).
\bibitem{kine}N.~Kinezaki, K.~Kawasaki, F.~Takasu and N.~Shigesada, {Theo. Popl. Biol.}, {\bf 64}, 291(2003).
\bibitem{hecht}I.~Hecht, Y.~Moran, and H.~Taitelbaum, {Phys. Rev. E}, {\bf 73}, 051109(2006).

\bibitem{havl}S.~Havlin and D.~ben Avraham, {Adv. Phys.} {\bf 36}, 695(1987).

\bibitem{balagurov}B.~Y. Balagurov and V.~G. Vaks, {Sov. Phys. JETP}, {\bf 38}, 968(1974).
\bibitem{donsker}M.~D. Donsker and S.~R.~S. Varadhan, {Commun. Pure Appl. Math.}, {\bf 32}, 721(1979).
\bibitem{gras}P.~Grassberger and I.~Procaccia, {J. Chem. Phys.}, {\bf 77}, 6281 (1982).
\bibitem{jayan1}P.~K.~Datta and A. M. Jayannavar, {Pramana-J. Phys.}, {\bf 38}, 257(1992).
\bibitem{wio}A.~D.~Sanchez, M.~Rodriguez and H.~S.~Wio, {Phys. Rev. E}, {\bf 57}, 6398(1998).
\bibitem{htait}H.~Taitelbaum, {Physica A}, {\bf 200}, 155(1993).
\bibitem{koza}H.~Taitelbaum, and Z.~Koza, {Physica A}, {\bf 285}, 166(2000).
\bibitem{wei89}G.~H.~Weiss, R.~Kopelman, and S~Havlin, {Phys. Rev. A}, {\bf 39}, 446 (1989).
\bibitem{tait}H.~Taitelbaum, R.~Kopelman, G~H Weiss, and S.~Havlin {Phys. Rev. A} {\bf 41}, 3116 (1990).

\bibitem{jayan2}P.~K.~Datta and A.~M.~Jayannavar, {Physica A}, {\bf 184}, 135 (1992).
\bibitem{thm}T.~M.~Nieuwenhuizen and H.~Brand, {J. Stat. Phys.}, {\bf 59}, 53 (1990).
\bibitem{jayan3}A.~M.~Jayannavar and J.~Kohler, {Phys. Rev A}, {\bf 41}, 3391(1990).
\bibitem{krpv}P.~L.~Krapivsky and K.~Mallick, {J. Stat. Mech.}, {\bf 2011}, P01015(2011).
\bibitem{tab}T.~Bagarti and K.~Kundu, {\it to be submitted}.
\end{thebibliography}
\end{document}